\documentclass[11pt]{article}

\usepackage[T1]{fontenc}
\usepackage[utf8]{inputenc}

\usepackage{graphicx}
\usepackage{amsmath,amssymb}
\usepackage{booktabs}
\usepackage{hyperref}
\usepackage{authblk}

\title{\textbf{crossfit: A Graph-Based Cross-Fitting Engine for Double/Debiased Machine Learning and Meta-Learners in R}}

\author[1]{Etienne Peyrot\footnote{ORCID: 0009-0006-8520-6201}}
\author[1]{Fran\c{c}ois Petit\footnote{ORCID: 0000-0003-2258-170X}}
\affil[1]{Universit\'e Paris Cit\'e and Universit\'e Sorbonne Paris Nord, Inserm, INRAE, Center for Research in Epidemiology and StatisticS (CRESS), F-75004 Paris, France}

\date{} % (set if needed)

\begin{document}
\maketitle

\begin{abstract}
Cross-fitting is a key ingredient in many semiparametric estimation procedures, such as double/debiased machine learning (DML), enabling valid estimation of low-dimensional targets in the presence of high-dimensional nuisance functions by enforcing out-of-sample use of nuisance predictions. \texttt{crossfit} is an R package that provides a general-purpose, estimator-agnostic cross-fitting engine. Users specify (i) a target functional and (ii) a directed acyclic graph (DAG) of nuisance models, with node-specific training fold widths and target-specific evaluation windows. The engine executes a reproducible schedule over folds, panels, and repetitions, returning either a scalar estimate (\texttt{mode="estimate"}) or a cross-fitted predictor function for application to new data (\texttt{mode="predict"}).

Beyond standard cross-fitting, \texttt{crossfit} implements fold-allocation modes that control how training data are shared across nuisance components, including disjoint and independence-enforcing allocations that duplicate reused nodes to reduce dependence between nuisance branches. The implementation targets simulation-heavy benchmarking and method development, with explicit and auditable schedules, defensive validation of specifications and nuisance dependencies, reuse-aware caching to avoid redundant refits, and failure isolation policies for large experiment grids. The \texttt{crossfit} package is available on CRAN, openly developed on GitHub under GPL-3, and is intended as a lightweight, tested foundation to prototype and empirically evaluate cross-fitted estimators with explicit control over fold geometry, dependence, and computation.
\end{abstract}


% ====================================================================
\section{Introduction}
\label{sec:introduction}

Cross-fitting is a sample-splitting method used to reduce bias and recover efficiency in estimators of parameters that depend on nonparametric or high-dimensional estimates of nuisance functions. When nuisance models are both trained and evaluated on the same data, the resulting correlation between the estimation and evaluation steps can introduce estimation errors.

The core idea behind cross-fitting is to use separate samples to estimate the nuisance functions and to evaluate them. Cross-fitting has a long history in semiparametric statistics and is widely used (see, for example, \cite{bickel1982,schick1986,zheng2011,newey2018,chernozhukov2018,foster2023}, among others). More recently, it has gained significant attention as it is a central component of Double/Debiased Machine Learning (DML) \cite{chernozhukov2018} and orthogonal learning \cite{foster2023}. By combining Neyman-orthogonal scores with cross-fitting, these approaches enable the use of machine learning methods to estimate nuisance parameters while still providing semiparametrically efficient and asymptotically normal estimators of the parameter of interest under relatively mild conditions.

The package \texttt{crossfit} was developed based on our experience conducting a large benchmark of estimators used in causal inference (Preliminary versions of \texttt{crossfit} were used in \cite{bouvier2024}). Studying the finite-sample properties of cross-fitted estimators requires large simulation experiments in which estimator performance is evaluated across multiple simulated scenarios. Because nuisance models must be repeatedly refitted in each realization, computation becomes a primary bottleneck. Moreover, cross-fitting implementations are susceptible to subtle errors, particularly data leakage and unintended in-sample reuse. For example, when preprocessing steps such as imputation, standardization, or feature selection are fit on the full dataset (rather than within each training fold) and then applied across folds.

Software support for cross-fitting is growing, but many tools focus on a fixed set of estimators and model classes. For example, the \texttt{DoubleML} (Python and R) ecosystem provides a high-level interface for canonical causal models and associated inference procedures \cite{bach2024}.

In contrast, \texttt{crossfit} was developed to support simulation-intensive methodological research and flexible implementation of cross-fitted estimators. The package \texttt{crossfit} is designed to (i) make the cross-fitting schedule explicit and auditable, (ii) support a broad variety of user-defined targets and nuisance learners, (iii) handle dependency structures between nuisance components through a DAG abstraction, and (iv) reduce duplicated computation when comparing multiple estimators under shared nuisance components.

This design supports both applied work (implementing cross-fitted estimators safely) and methodological work (systematically studying empirical properties of cross-fitting/double debiasing under different fold geometries, repetitions, and learning strategies).

\paragraph{Statement of need.}
Cross-fitting is a standard ingredient in modern semiparametric and causal estimation, nonetheless its empirical behavior (finite-sample stability, sensitivity to fold geometry, number of folds, dependence between nuisance learners, interaction with nuisance complexity) remains an active research topic \cite{okasa2022,ellul2025}. A practical barrier to empirical study is that cross-fitting implementations are easy to "get almost right" while silently introducing leakage (e.g., using nuisance predictions trained on the same observations used in target evaluation, or fitting preprocessing steps, such as imputation or standardization, on the full dataset instead of within-fold training data), inconsistent fold schedules across methods, unintended dependence structures between nuisance learners, and fold-specific convergence or fitting failures of nuisance learners (which can occur in large simulation studies and can propagate to the final estimate).

\texttt{crossfit} addresses this need by providing:
\begin{itemize}
  \item A \textbf{general interface} for defining nuisance learners and targets without committing to a particular estimator family.
  \item A \textbf{graph-based abstraction} for nuisance dependencies, enabling complex multi-stage pipelines (nuisance-of-nuisance constructions) to be represented explicitly.
  \item \textbf{Explicit control} over fold geometry: node-specific training fold width (\texttt{train\_fold}), target evaluation window width (\texttt{eval\_fold}), number of folds \texttt{K}, and repetition count.
  \item \textbf{Well-defined fold-allocation modes} that govern how training folds are assigned across nuisance instances and, crucially, whether dependence is reduced by allocating disjoint training folds and/or duplicating shared nodes (Figure~\ref{fig:fold-allocation}).
  \item \textbf{Defensive validation} (argument checks, dependency coverage checks, cycle detection) and \textbf{automated tests} that verify cross-fitting invariants such as the absence of leakage.
  \item \textbf{Computation-aware execution} intended for large simulation studies: structured execution over the nuisance graph and reuse/caching of identical nuisance work across methods and panels when possible.
\end{itemize}

% ====================================================================
\section{Package overview and design}
\label{sec:design}

This section describes the main user-facing abstractions and the design choices that structure \texttt{crossfit}.

\subsection{User-facing abstractions}
\label{subsec:user-facing}

\texttt{crossfit} is built around two user-facing building blocks.

\begin{itemize}
  \item \textbf{\texttt{create\_nuisance()}} defines a nuisance node by providing a \texttt{fit()} function and a \texttt{predict()} function. The interface is learner-agnostic: \texttt{fit()} may call anything from \texttt{lm()} to external machine learning libraries, as long as it returns a model object that \texttt{predict()} can consume. A nuisance node can specify \texttt{train\_fold}, the number of folds used for training at that node (supporting designs where different nuisance components use different training widths).
  \item \textbf{\texttt{create\_method()}} defines a target functional together with the required nuisances and cross-fitting parameters: number of folds \texttt{K}, number of repetitions, target evaluation window width \texttt{eval\_fold}, a fold allocation mode, and aggregation functions over panels and repetitions. Targets are standard R functions written with explicit named nuisance arguments, e.g. \texttt{target(data, nuis\_y, nuis\_m, ...)}.
\end{itemize}

Two modes are supported:
\begin{itemize}
  \item \textbf{\texttt{mode="estimate"}}: the engine returns a numeric estimate of the target.
  \item \textbf{\texttt{mode="predict"}}: the engine returns a cross-fitted predictor function that can be applied to new data. This is useful when the target itself is a meta-learner that outputs predictions (e.g., a cross-fitted regression function, a T-learner/S-learner style predictor, or any user-defined mapping built from nuisance predictions). In this mode, aggregation functions combine panel-wise predictors into an ensemble predictor (e.g., mean/median across panels and repetitions).
\end{itemize}

\subsection{Dependency graph and validation}
\label{subsec:dag-validation}

A core design choice is to treat nuisance components as nodes in a directed acyclic graph (DAG). Dependencies arise when a nuisance's \texttt{fit()} or \texttt{predict()} function requires as input the predictions of other nuisance components. This also covers learned preprocessing steps, such as data imputation, normalization, calibration, or feature construction. Before execution, \texttt{crossfit} validates nuisance graphs and method specifications:
\begin{itemize}
  \item \textbf{Argument coverage checks:} required arguments in nuisance \texttt{fit()} / \texttt{predict()} must have corresponding nuisance mappings.
  \item \textbf{Cycle detection:} cyclic dependency graphs are rejected early with informative errors.
  \item \textbf{Target--nuisance consistency:} the target function's required nuisance arguments must be provided by the method specification.
  \item \textbf{Input sanitization:} fold parameter constraints and mode-specific requirements are enforced at specification time (early error detection).
\end{itemize}

These checks are especially important because many cross-fitting mistakes otherwise manifest as subtle leakage or silent in-sample reuse.

\subsection{Cross-fitting schedule: folds, repetitions, and panels}
\label{subsec:schedule}

For each method and repetition, the data are partitioned into \texttt{K} folds. Within a repetition, \texttt{crossfit} iterates over \textbf{panels}. A panel is one "position" of the evaluation window: for example, \texttt{eval\_fold = 1} evaluates the target on a single fold, while larger \texttt{eval\_fold} evaluates on a window of consecutive folds. Across panels, the evaluation window advances by a cyclic shift (step size 1 modulo \texttt{K}). At each panel, nuisances are trained on their assigned training folds, nuisance predictions are computed on the target evaluation fold/window, and the target is evaluated using those out-of-sample predictions.

\subsection{Fold allocation modes}
\label{subsec:allocation}

All fold allocation modes in \texttt{crossfit} satisfy the same central requirement: in every mode, the target evaluation fold/window is disjoint from all nuisance training folds used to generate the nuisance predictions passed to the target. This is the usual out-of-sample guarantee targeted by standard DML cross-fitting \cite{chernozhukov2018,newey2018}.

\paragraph{Allocation of training folds across nuisances.}
Modes differ only in how training folds are assigned across nuisance instances inside a panel, i.e., whether different nuisance learners (or different branches of a nuisance dependency graph) are allowed to reuse the same training folds.

This distinction matters because, beyond the leading orthogonality-driven cancellation, many doubly robust / orthogonal estimators have a remainder containing products of nuisance estimation errors. If nuisance errors are correlated because the nuisances were trained on the same data, this "nonlinear" term can behave poorly in finite samples. Several papers therefore consider strengthened splitting geometries that deliberately separate training samples across nuisance components. The terminology varies across the literature ("double cross-fitting" \cite{newey2018,mcclean2024}, "three-way cross-fitting" \cite{fisher2023}, "triple cross-fit" \cite{ke2024}, and "multiway cross-fitting" \cite{chiang2022}) but the common principle is to train the nuisance components that interact in the remainder term on separate subsamples.

\begin{figure}[ht]
  \centering
  \includegraphics[width=\linewidth]{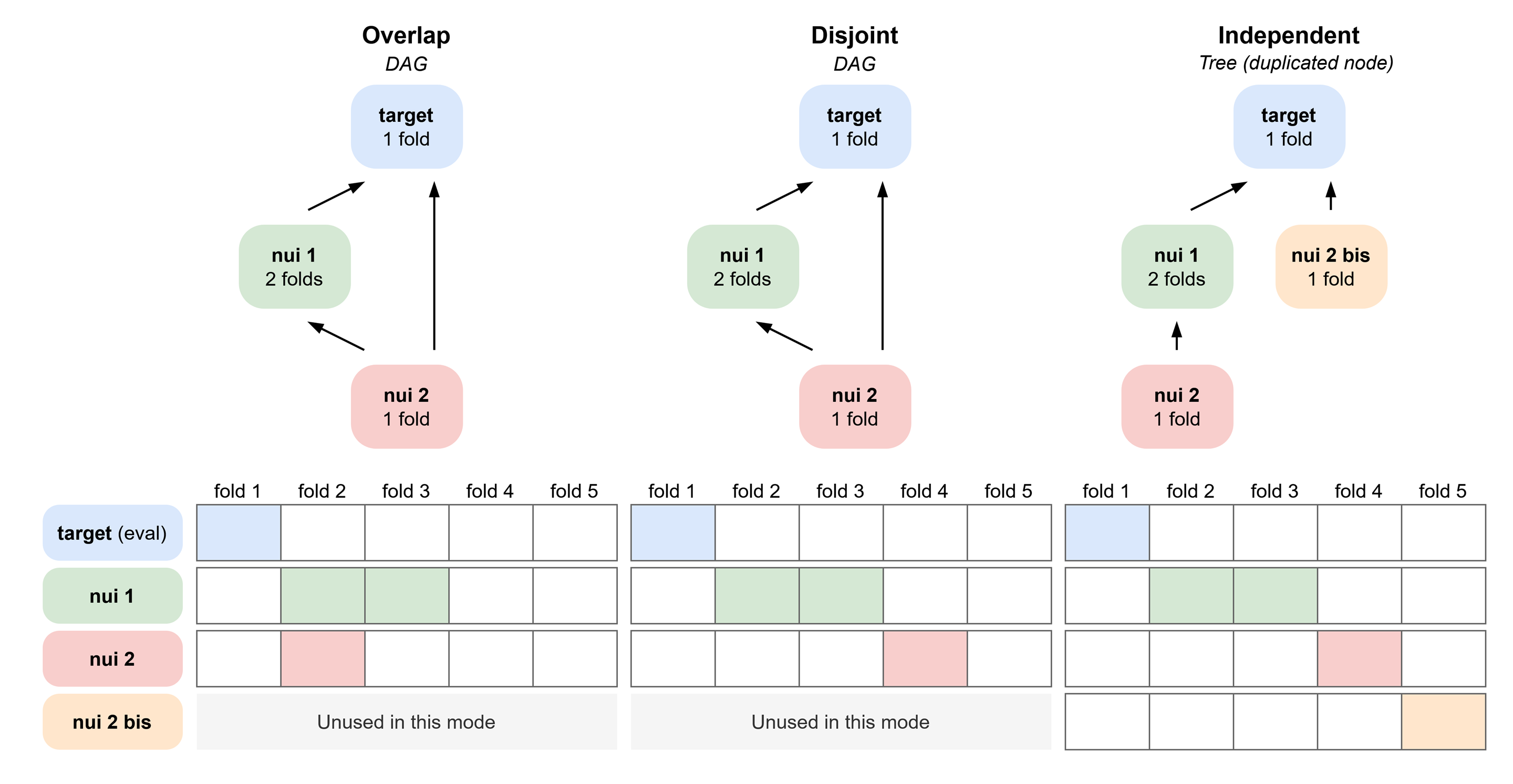}
  \caption{\textbf{Fold allocation modes in \texttt{crossfit}.} The same method is executed under three fold allocation modes. \texttt{target} denotes the user-defined target functional (the final quantity computed), and \texttt{nui 1}/\texttt{nui 2} denote two required nuisance models; the arrow \texttt{nui 2 $\rightarrow$ nui 1} indicates a nuisance-of-nuisance dependency. Top: nuisance dependency structure and node-specific training widths (\texttt{train\_fold} = number of folds used to train each nuisance). Bottom: for \texttt{K=5}, the panel where \texttt{target} is evaluated on fold 1 (blue) is shown. Colored cells indicate the training folds used to fit each nuisance instance in that panel. \texttt{nui 2 bis} is a duplicated copy of \texttt{nui 2} used only in Independent mode.}
  \label{fig:fold-allocation}
\end{figure}

\paragraph{How the three modes map to these ideas.}
Figure~\ref{fig:fold-allocation} shows the same method (a triangle-shaped nuisance DAG) executed under the three modes for \texttt{K=5} and for the single panel where the target is evaluated on fold 1 (blue). Other panels are identical up to a cyclic shift modulo \texttt{K}.

Concretely, in \texttt{crossfit}:
\begin{itemize}
  \item \textbf{\texttt{"overlap"}} corresponds to the original, widely used cross-fitting geometry in DML: it enforces the out-of-sample constraint with respect to the evaluation fold/window, but allows different nuisances to be trained on overlapping (often identical) training folds within a panel \cite{chernozhukov2018,newey2018}.
  \item \textbf{\texttt{"disjoint"}} implements the core idea behind "double", "triple", "three-way" designs in their simplest form: within a panel, it coordinates training-fold assignment across nuisance instances to avoid fold reuse when possible, so nuisance models that jointly contribute to a second-order remainder term can be trained on different data \cite{mcclean2024,fisher2023,ke2024}. It does not duplicate shared nodes, so when a nuisance is reused (directly or indirectly) in multiple places, some dependence can remain through shared upstream fits.
  \item \textbf{\texttt{"independence"}} enforces a stronger version of the same principle: if a nuisance node is reused across branches, the engine duplicates that node (tree expansion) so that each branch can be trained on disjoint folds, aiming to remove dependence between branches in the sense targeted by multiway cross-fitting constructions \cite{chiang2022}. This can require a larger \texttt{K} (more available folds) and can increase the number of nuisance fits, but yields the cleanest separation when a method reuses nuisances in multiple places.
\end{itemize}

\paragraph{Practical trade-offs.}
Overlap mode corresponds to the default geometry in classical DML and is often a natural baseline \cite{chernozhukov2018}. Disjoint and independence modes allow users to control how much training data is shared across nuisance learners, which connects to an active research question: when do stronger splitting geometries (double, triple, three-way, multiway cross-fitting) yield better finite-sample performance by limiting dependence-driven second-order bias terms \cite{mcclean2024,fisher2023,ke2024,chiang2022,zivich2021}? \texttt{crossfit} is designed to make such comparisons straightforward.

\paragraph{Computation and fold allocation.}
Both \texttt{"disjoint"} and \texttt{"independence"} may increase the number of nuisance instances that appear in a panel (because folds must be separated, and because \texttt{"independence"} duplicates reused nodes). However, within a repetition the panel schedule is cyclic (each panel is a shift modulo \texttt{K}), so the set of training-fold windows requested by a given nuisance specification repeats across panels up to permutation. In \texttt{crossfit}, reusable nuisance fits are cached within each repetition (see the caching description below), so the additional work induced by stronger allocation modes is often front-loaded in early panels (cache warm-up) and reused in later panels. When \texttt{"independence"} requires a larger minimal \texttt{K}, any added computational cost primarily comes from the increased number of panels rather than repeated refitting of identical nuisance specifications on identical fold windows.

\subsection{Performance and computational considerations}
\label{subsec:performance}

A motivating use case for \texttt{crossfit} is simulation-heavy methodological work (e.g., stress-testing cross-fitted estimators across fold allocation modes, learner choices, sample sizes, and repetitions). In this setting, performance is primarily driven by nuisance fitting, and \texttt{crossfit} aims to keep cross-fitting overhead small by (i) executing computations in dependency order over the nuisance graph, (ii) running multiple methods under shared schedules when possible, and (iii) caching fitted nuisances only if reused (within a repetition) so that cyclic panel shifts amortize repeated requests for the same training-fold windows.

\begin{itemize}
  \item \textbf{Graph-structured execution:} nuisance evaluation respects the dependency structure (DAG or tree-expanded graph), yielding a transparent execution plan and avoiding redundant upstream computations when multiple downstream quantities depend on the same nuisance.
  \item \textbf{Cross-method reuse with \texttt{crossfit\_multi()}:} researchers often compare several targets under identical splits. \texttt{crossfit\_multi()} runs multiple methods in a single call, enabling shared scheduling and facilitating reuse of shared nuisance components across methods.
  \item \textbf{Instance normalization and reuse-aware caching:} internally, nuisance requests are compiled into normalized instances identified by (i) a structural signature of the nuisance specification and dependency context and (ii) the exact training-fold window used in a panel. When two requests correspond to the same instance, the engine reuses the fitted object rather than refitting. To keep memory and overhead under control, \texttt{crossfit} does selective caching within each repetition.
  \item \textbf{Allocation-aware compute trade-offs:} \texttt{"disjoint"} and especially \texttt{"independence"} can increase the number of nuisance nodes appearing in a panel (due to constrained fold assignment and, for \texttt{"independence"}, tree expansion by duplicating reused nodes). In \texttt{crossfit}, this does not automatically translate into proportionally more fitting work because the panel schedule cycles over \texttt{K} shifts.
  \item \textbf{Failure control for large experiments:} nuisance learners may fail on some folds (non-convergence, numerical issues). \texttt{crossfit} records failures and supports a \texttt{max\_fail} policy to stop spending compute on repeatedly failing methods while allowing other methods to continue.
  \item \textbf{Deterministic fold splitting for reproducible benchmarking:} users may supply a deterministic \texttt{fold\_split} function, enabling exact reproduction of schedules across machines and code revisions.
\end{itemize}

% ====================================================================
\section{Typical workflow}
\label{sec:workflow}

\subsection{Estimation mode}
Define a nuisance learner, define a target functional, build a method specification, then run cross-fitting.

\begin{verbatim}
nuis_y <- create_nuisance(
  fit = function(data) {
    ranger::ranger(y ~ x, data = data, num.trees = 500, seed = 1)
  },
  predict = function(model, data) {
    as.numeric(predict(model, data = data)$predictions)
  }
)

target_mse <- function(data, nuis_y, ...) mean((data$y - nuis_y)^2)

m <- create_method(
  target = target_mse,
  list_nuisance = list(nuis_y = nuis_y),
  folds = 2,
  repeats = 2,
  eval_fold = 1L,
  mode = "estimate",
  aggregate_panels = mean_estimate,
  aggregate_repeats = median_estimate,
  fold_allocation = "independence"
)

res <- crossfit(data, m)
res$estimates
\end{verbatim}

\subsection{Prediction mode}
In \texttt{mode="predict"}, the target returns a vector of predictions (on the evaluation window); \texttt{crossfit} aggregates panel-wise predictors into a single predictor function for new data.

\begin{verbatim}
target_pred <- function(data, nuis_y) nuis_y

m_pred <- create_method(
  target = target_pred,
  list_nuisance = list(nuis_y = nuis_y),
  folds = 2,
  repeats = 2,
  eval_fold = 0L,
  mode = "predict",
  aggregate_panels = mean_predictor,
  aggregate_repeats = median_predictor,
  fold_allocation = "independence"
)

res_pred <- crossfit(data, m_pred)

pred_fun <- res_pred$estimates$pred
pred_fun(newdata)
\end{verbatim}

% ====================================================================
\section{Technical details and quality control}
\label{sec:technical}

\subsection{Automated unit and integration testing}
\texttt{crossfit} includes automated tests implemented with \texttt{testthat} (edition 3). Unit tests cover specification helpers (\texttt{create\_nuisance()}, \texttt{create\_method()}), aggregator utilities (including predictor aggregators), and validation logic (e.g., missing or mismatched nuisance arguments). Integration tests cover the main execution functions (\texttt{crossfit()}, \texttt{crossfit\_multi()}) on small synthetic problems and verify expected output structure (estimates, names, repetition counters) in both estimation and prediction modes.

\subsection{Cross-fitting invariants: no-leakage testing}
Because the central correctness requirement of cross-fitting is that target evaluation uses nuisance predictions trained out-of-sample, the test suite includes a deterministic, trace-based invariant test. In this test, fold membership is embedded directly into the data and a deterministic fold splitter is used. Nuisance \texttt{fit()} and \texttt{predict()} functions return structured trace strings encoding which folds were used for training and which folds were predicted. The target function parses these traces and asserts that the evaluation fold never appears among the training folds of the nuisance models whose predictions it receives. This invariant is checked across supported fold allocation modes (\texttt{"overlap"}, \texttt{"disjoint"}, \texttt{"independence"}), providing a strong regression test against accidental leakage introduced by future refactoring.

\subsection{Robustness: failure isolation}
Cross-fitting pipelines may include nuisance learners that fail on some folds or repetitions (e.g., due to convergence issues or data-dependent errors). \texttt{crossfit} isolates such failures so that one failing method does not crash other methods executed in the same call. The test suite includes an integration test where one method is constructed to fail deterministically during nuisance fitting; the test verifies that:
\begin{itemize}
  \item a non-failing method still returns a valid estimate,
  \item the failing method records errors and returns \texttt{NA} when no repetitions succeed,
  \item the realized number of successful repetitions is tracked per method (useful for diagnosing instability and tuning \texttt{max\_fail}).
\end{itemize}

\subsection{Continuous integration}
Continuous integration is implemented via GitHub Actions. On every push and pull request, the workflow runs \texttt{R CMD check} (with vignettes built and the PDF manual disabled via \texttt{--no-manual}) across a matrix of operating systems and R versions, including macOS (R release), Windows (R release), and Ubuntu Linux (R release and R devel). This automated matrix reduces OS-specific regressions and helps ensure forward compatibility with upcoming R releases.

\section{Discussion, limitations, and reuse}
\label{sec:discussion}

\subsection{Limitations and scope}
The \texttt{crossfit} package is a cross-fitting engine rather than a library of ready-made estimators. As a consequence, several limitations follow.
\begin{itemize}
  \item \textbf{Estimator-agnostic core:} the package does not provide built-in causal estimators, orthogonal scores, or standard-error formulas. Users implement targets and inference procedures externally and call \texttt{crossfit} to obtain cross-fitted nuisance predictions and/or target evaluations.
  \item \textbf{Fold-geometry constraints:} stronger allocation modes such as \texttt{"disjoint"} and especially \texttt{"independence"} may require larger numbers of folds \texttt{K} (or may be infeasible for a given nuisance graph and node-specific \texttt{train\_fold} settings). They can also increase the number of nuisance instances to fit, trading statistical dependence reduction for additional computation.
  \item \textbf{No built-in parallelization:} the engine is deterministic and reproducible but does not internally parallelize over folds/panels/repetitions. Users can parallelize at a higher level (e.g., over repetitions or simulation replications) using standard R tooling.
  \item \textbf{Correctness depends on the user-defined interface:} \texttt{crossfit} enforces out-of-sample fold separation in its schedule, but user-defined nuisance code should avoid unintended side effects (e.g., using global state, caching data across calls, or accessing outcome variables in predictors).
\end{itemize}

\subsection{Reuse potential}
The \texttt{crossfit} package can be reused wherever researchers need to enforce out-of-sample evaluation of learned nuisance functions, including learned preprocessing steps (e.g., imputation), when estimating a low-dimensional target. Typical use cases include DML estimators for causal parameters (e.g., average treatment effects, partially linear models), orthogonal-score-based targets in semiparametric estimation, and meta-learners that require multiple nuisance components with dependencies. Because nuisances are defined through user-provided \texttt{fit()} and \texttt{predict()} functions, \texttt{crossfit} can wrap a wide range of learners available in R without imposing a particular modeling framework.

\paragraph{Prediction reuse.}
In addition to scalar target estimation, \texttt{mode="predict"} supports workflows where the object of interest is a predictor constructed from cross-fitted nuisance predictions. This can be reused to (i) generate out-of-sample predictions for downstream evaluation, (ii) build meta-learners that require cross-fitting (e.g., plug-in predictors built from multiple nuisances), or (iii) export a stable prediction function for application to new datasets while preserving the cross-fitting discipline used during training.

\paragraph{Empirical methodology reuse.}
The package is well-suited to simulation and benchmarking studies: users can vary \texttt{K}, repetitions, node-specific training widths, and fold allocation modes to study finite-sample behavior and stability under controlled schedules, including dependence-reduction geometries motivated by double, triple, and three-way cross-fitting ideas \cite{mcclean2024,fisher2023,ke2024,chiang2022}. The deterministic \texttt{fold\_split} interface supports reproducible experiments, including structured splits (grouped folds, time-based folds) when appropriate. \texttt{crossfit\_multi()} encourages fair comparisons by enforcing identical splits across competing estimators and by enabling shared nuisances to be reused across methods.

\subsection{Extending the software}
The core engine can be extended in several directions:
\begin{itemize}
  \item \textbf{New targets:} implement new target functionals by writing an R function that consumes \texttt{data} and named nuisance predictions.
  \item \textbf{New nuisances:} wrap any learner with a fit/predict interface as a nuisance node; express multi-stage pipelines through nuisance dependencies.
  \item \textbf{Custom fold splitting:} supply a custom \texttt{fold\_split} function to define folds deterministically or under complex sampling schemes.
  \item \textbf{Aggregation and diagnostics:} provide custom aggregation functions over panels and repetitions, including predictor aggregation rules for \texttt{mode="predict"}.
\end{itemize}

\subsection{Support and contributions}
Development and support are managed through the GitHub repository. Users can file bug reports and feature requests through the issue tracker (\texttt{/issues}). Contributions are welcomed via pull requests; continuous integration provides automated feedback on portability and correctness across the supported OS/R matrix.

% ====================================================================
\section*{Availability}
\label{sec:availability}

\paragraph{Operating system.} Any operating system supported by R. Automated checks run on Ubuntu Linux, Windows, and macOS via GitHub Actions.

\paragraph{Programming language.} R ($\geq 4.1.0$).

\paragraph{System requirements.} None beyond a standard R installation.

\paragraph{Dependencies.} Imports: \texttt{stats}, \texttt{utils}.

\paragraph{Source code and archive.}
\paragraph{Source code and archive.}
The \texttt{crossfit} package is available on CRAN: \url{https://CRAN.R-project.org/package=crossfit}. Development occurs on GitHub under GPL-3: \url{https://github.com/EtiennePeyrot/crossfit-R}.

\paragraph{Installation.}
\begin{verbatim}
\begin{verbatim}
# Install the released version from CRAN
install.packages("crossfit")

# Install the development version from GitHub
install.packages("remotes")
remotes::install_github("EtiennePeyrot/crossfit-R", build_vignettes = TRUE)
\end{verbatim}

% ====================================================================
\section*{Acknowledgements}
The author thanks the open-source R community for foundational tooling (including \texttt{testthat}, \texttt{knitr}, and \texttt{rmarkdown}) that supports reliable package development and documentation.

\section*{Funding}
Etienne Peyrot acknowledges support from the Universit\'e Paris Cit\'e. Francois Petit acknowledges support from the French Agence Nationale de la Recherche through the project reference ANR-22-CPJ1-0047-01.

\section*{Competing interests}
The author declares that they have no competing interests.

\appendix

\section{Recipe: DML for the partially linear regression model}
\label{app:plr-pension}

This appendix gives a complete, self-contained example of a Double/Debiased Machine Learning (DML) estimator for the Partially Linear Regression (PLR) model using the \texttt{pension} dataset shipped with the \texttt{hdm} package. The example illustrates how \texttt{crossfit} can be used as an estimator-agnostic cross-fitting engine: the user supplies nuisance learners and a target functional, and \texttt{crossfit} executes an auditable, out-of-sample schedule. This example follows the classical 401(k) application in \cite{chernozhukov2018}.

\subsection{Model and estimand}

In the PLR model, the effect of a treatment variable $D$ on an outcome $Y$ is assumed to be linear with coefficient $\theta_0$, while the effect of covariates $X$ is allowed to be nonparametric:
\begin{align}
  Y &= D \theta_0 + g_0(X) + U,\qquad \mathbb{E}\left[U \mid X, D\right] = 0,\\
  D &= m_0(X) + V,\qquad \mathbb{E}\left[V \mid X\right] = 0.
\end{align}
When $D$ is binary, $m_0(X) = \mathbb{E}\left[D \mid X\right] = \mathbb{P}\left(D=1 \mid X\right)$ is the propensity score. DML estimates $\theta_0$ by (i) estimating nuisance functions $g_0$ and $m_0$ with flexible learners, (ii) residualizing $Y$ and $D$, and (iii) regressing the residualized outcome on the residualized treatment using \emph{cross-fitted} nuisance predictions (trained out-of-sample).

\subsection{Data: the \texttt{pension} 401(k) dataset}

We use \texttt{pension} from \texttt{hdm}. Following common practice, we take:
\begin{itemize}
  \item outcome $Y =$ \texttt{tw} (total wealth),
  \item treatment $D =$ \texttt{p401} (participation in a 401(k), binary),
  \item controls $X =$ a standard subset of demographic and income indicators (listed below).
\end{itemize}

\subsection{Implementation with \texttt{crossfit}}

The DML "partialling-out" estimator can be written as:
\[
  \widehat{\theta}
  \;=\;
  \frac{\sum_{i \in \mathcal{I}_{\mathrm{eval}}} \widetilde{D}_i \widetilde{Y}_i}{\sum_{i \in \mathcal{I}_{\mathrm{eval}}} \widetilde{D}_i^2},
  \qquad
  \widetilde{Y}_i = Y_i - \widehat{g}(X_i),\;
  \widetilde{D}_i = D_i - \widehat{m}(X_i),
\]
where $\widehat{g}$ and $\widehat{m}$ are trained on folds disjoint from the evaluation set $\mathcal{I}_{\mathrm{eval}}$ (cross-fitting). We implement this by defining two nuisance learners and a target functional in \texttt{crossfit}.

\subsubsection*{Complete code}

This appendix uses the suggested packages \texttt{hdm}, \texttt{ranger}, and \texttt{glmnet} for the examples; they are not required dependencies of \texttt{crossfit}.
\begin{verbatim}
library(hdm)
library(ranger)
library(crossfit)

data(pension)

y <- pension$tw
d <- pension$p401

xvar <- c(
  "i2","i3","i4","i5","i6","i7",
  "a2","a3","a4","a5",
  "fsize","hs","smcol","col",
  "marr","twoearn","db","pira","hown"
)

X <- pension[ , xvar]
dat <- data.frame(y = y, d = d, X)
dat <- na.omit(dat)

nuis_g <- create_nuisance(
  fit = function(data) {
    ranger::ranger(y ~ . - d, data = data, num.trees = 20, seed = 1)
  },
  predict = function(model, data) {
    as.numeric(predict(model, data = data)$predictions)
  }
)

nuis_m_rf <- create_nuisance(
  fit = function(data) {
    data$d <- factor(data$d, levels = c(0, 1))   # ensure classification
    ranger::ranger(d ~ . - y, data = data, probability = TRUE,
                   num.trees = 20, seed = 1)
  },
  predict = function(model, data) {
    as.numeric(predict(model, data = data)$predictions[ , "1"])
  }
)

target_plr <- function(data, nuis_g, nuis_m) {
  y_tilde <- data$y - nuis_g
  d_tilde <- data$d - nuis_m
  sum(d_tilde * y_tilde) / sum(d_tilde^2)
}

m_plr <- create_method(
  target = target_plr,
  list_nuisance = list(nuis_g = nuis_g, nuis_m = nuis_m_rf),
  folds = 5,
  repeats = 2,
  eval_fold = 1L,
  mode = "estimate",
  fold_allocation = "overlap",
  aggregate_panels = mean_estimate,
  aggregate_repeats = median_estimate
)

res_plr <- crossfit(dat, m_plr)
res_plr$estimates
\end{verbatim}

\begin{verbatim}
#> $method_1
#> [1] 11487.8
\end{verbatim}

\subsection{Fold allocation modes}
A central feature of \texttt{crossfit} is that the same estimator can be run under different fold-allocation policies (\texttt{"overlap"}, \texttt{"disjoint"}, \texttt{"independence"}) without changing nuisance or target definitions.

\begin{verbatim}
m_plr_overlap <- m_plr

m_plr_disjoint <- create_method(
  target = target_plr,
  list_nuisance = list(nuis_g = nuis_g, nuis_m = nuis_m_rf),
  folds = 5,
  repeats = 2,
  eval_fold = 1L,
  mode = "estimate",
  fold_allocation = "disjoint",
  aggregate_panels = mean_estimate,
  aggregate_repeats = median_estimate
)

m_plr_indep <- create_method(
  target = target_plr,
  list_nuisance = list(nuis_g = nuis_g, nuis_m = nuis_m_rf),
  folds = 5,
  repeats = 2,
  eval_fold = 1L,
  mode = "estimate",
  fold_allocation = "independence",
  aggregate_panels = mean_estimate,
  aggregate_repeats = median_estimate
)

res_modes <- crossfit_multi(
  data = dat,
  methods = list(
    overlap = m_plr_overlap,
    disjoint = m_plr_disjoint,
    independence = m_plr_indep
  )
)

res_modes$estimates
\end{verbatim}

\begin{verbatim}
#> $overlap
#> [1] 11248.5
#> 
#> $disjoint
#> [1] 10599.07
#> 
#> $independence
#> [1] 10599.07
\end{verbatim}

\paragraph{Notes.} (i) The nuisance learners used here (linear and logistic regression) are chosen for portability and speed. In typical DML applications, they can be replaced by more flexible learners without changing the \texttt{crossfit} interface. (ii) \texttt{"disjoint"} and especially \texttt{"independence"} may require larger $K$ for some nuisance graphs and training-fold widths; if infeasible, \texttt{crossfit} reports a specification error early during validation.

\section{Recipe: Nuisance dependency graphs (nuisance-of-nuisance)}
\label{app:dag}

(i) We use random forests (package \texttt{ranger}) to illustrate nonparametric nuisance learning. For each random-forest we set \texttt{num.trees\ =\ 20} for speed in the article; in applications you typically increase it. Later, we additionally compare a random-forest propensity score with an \(\ell_1\)-penalized logistic regression (package \texttt{glmnet}). Any learner with a \texttt{fit()}/\texttt{predict()} interface can be plugged into \texttt{crossfit} without changing the target functional. (ii) \texttt{"disjoint"} and especially \texttt{"independence"} may require larger \(K\) for some nuisance graphs and training-fold widths; if infeasible, \texttt{crossfit} reports a specification error early during validation.

\subsection{Propensity-augmented outcome regression inside PLR}

The code below defines the propensity nuisance \texttt{nuis\_m}, then defines an outcome nuisance \texttt{nuis\_g\_ps} whose \texttt{fit()} and \texttt{predict()} functions take \texttt{nuis\_m} as an additional argument. This induces the edge \texttt{nuis\_m $\rightarrow$ nuis\_g\_ps} in the nuisance graph. The target remains the PLR partialling-out estimator.

\begin{verbatim}
nuis_g_ps <- create_nuisance(
  fit = function(data, nuis_m) {
    data$nuis_m <- nuis_m
    ranger::ranger(y ~ nuis_m + . - d, data = data, num.trees = 20, seed = 1)
  },
  predict = function(model, data, nuis_m) {
    data$nuis_m <- nuis_m
    as.numeric(predict(model, data = data)$predictions)
  }
)
m_plr_dag <- create_method(
  target = target_plr,
  list_nuisance = list(nuis_g = nuis_g_ps, nuis_m = nuis_m_rf),
  folds = 5,
  repeats = 2,
  eval_fold = 1L,
  mode = "estimate",
  fold_allocation = "overlap",
  aggregate_panels = mean_estimate,
  aggregate_repeats = median_estimate
)

res_plr_dag <- crossfit(dat, m_plr_dag)
res_plr_dag$estimates
\end{verbatim}

\begin{verbatim}
#> $method_1
#> [1] 13169.04
\end{verbatim}

\paragraph{Remark.} The goal here is to illustrate the dependency mechanism. In practice, nuisance-of-nuisance constructions can encode multi-stage ML pipelines, calibration steps, imputation steps, stacked learners, or other intermediate quantities that must be computed out-of-sample.

\section{Recipe: Prediction mode (returning a cross-fitted predictor function)}
\label{app:predict}

In addition to scalar estimation, \texttt{crossfit} supports \texttt{mode="predict"}, where the target returns a predictor object and the engine aggregates panel-wise predictors into a single predictor function for application to new data. This is useful for meta-learners and for workflows where the primary output is a reusable prediction rule rather than a single scalar.

\subsection{Cross-fitted propensity score predictor}

The following recipe returns a cross-fitted predictor function for the propensity score $m_0(X) = \mathbb{E}\left[D \mid X\right]$ on the \texttt{pension} dataset.

\begin{verbatim}
target_ps <- function(data, nuis_m, ...) nuis_m

m_ps <- create_method(
  target = target_ps,
  list_nuisance = list(nuis_m = nuis_m_rf),
  folds = 5,
  repeats = 2,
  eval_fold = 0L,
  mode = "predict",
  fold_allocation = "overlap",
  aggregate_panels = mean_predictor,
  aggregate_repeats = median_predictor
)

res_ps <- crossfit(dat, m_ps)

pscore_fun <- res_ps$estimates$method_1
pscore_fun(dat[1:5, ])
\end{verbatim}

\begin{verbatim}
#> [1] 0.1743680 0.2398349 0.4336596 0.2087443 0.2807850
\end{verbatim}

\paragraph{Remark.} Although this example returns a propensity score predictor, the same pattern applies to any meta-learner that constructs predictions from cross-fitted nuisance outputs.

\section{Recipe: Benchmarking multiple methods with \texttt{crossfit\_multi()}}
\label{app:multi}

Simulation studies and method comparisons often require running several estimators under identical splits. The \texttt{crossfit\_multi()} function runs multiple method specifications in a single call, enabling shared scheduling and reuse of shared nuisance components when possible.

\subsection{Two PLR estimators with different propensity learners}

This example compares two PLR estimators that share the same outcome nuisance \texttt{nuis\_g} but use different propensity nuisances: a random-forest propensity (\texttt{ranger}) versus an $\ell_1$-penalized logistic regression (\texttt{glmnet}). The target functional \texttt{target\_plr} is reused unchanged.

\begin{verbatim}
nuis_m_lasso <- create_nuisance(
  fit = function(data) {
    x <- model.matrix(d ~ . - y - 1, data = data)
    y <- data$d
    glmnet::cv.glmnet(x, y, family = "binomial", alpha = 1)
  },
  predict = function(model, data) {
    x <- model.matrix(d ~ . - y - 1, data = data)
    as.numeric(predict(model, newx = x, s = "lambda.1se", type = "response"))
  }
)

m_plr_rf <- create_method(
  target = target_plr,
  list_nuisance = list(nuis_g = nuis_g, nuis_m = nuis_m_rf),
  folds = 5,
  repeats = 2,
  eval_fold = 1L,
  mode = "estimate",
  fold_allocation = "overlap",
  aggregate_panels = mean_estimate,
  aggregate_repeats = median_estimate
)

m_plr_lasso <- create_method(
  target = target_plr,
  list_nuisance = list(nuis_g = nuis_g, nuis_m = nuis_m_lasso),
  folds = 5,
  repeats = 2,
  eval_fold = 1L,
  mode = "estimate",
  fold_allocation = "overlap",
  aggregate_panels = mean_estimate,
  aggregate_repeats = median_estimate
)

res_cmp <- crossfit_multi(
  data = dat,
  methods = list(plr_rf = m_plr_rf, plr_lasso = m_plr_lasso)
)

res_cmp$estimates
\end{verbatim}

\begin{verbatim}
#> $plr_rf
#> [1] 10619.41
#> 
#> $plr_lasso
#> [1] 11115.44
\end{verbatim}

\paragraph{Remark.} This pattern generalizes to larger benchmarking grids. Because \texttt{crossfit\_multi()} enforces a shared fold schedule across methods within a call, it simplifies fair comparisons and supports reuse of shared nuisance work when method specifications overlap.


\begin{thebibliography}{10}

\bibitem{bickel1982}
Bickel PJ.
\newblock {On Adaptive Estimation}.
\newblock The Annals of Statistics. 1982;10(3):647-71.
\newblock Available from: \url{https://doi.org/10.1214/aos/1176345863}.

\bibitem{schick1986}
Schick A.
\newblock {On Asymptotically Efficient Estimation in Semiparametric Models}.
\newblock The Annals of Statistics. 1986;14(3):1139-51.
\newblock Available from: \url{https://doi.org/10.1214/aos/1176350055}.

\bibitem{zheng2011}
Zheng W, van~der Laan MJ.
\newblock Cross-Validated Targeted Minimum-Loss-Based Estimation.
\newblock In: Targeted Learning: Causal Inference for Observational and Experimental Data. New York, NY: Springer New York; 2011. p. 459-74.
\newblock Available from: \url{https://doi.org/10.1007/978-1-4419-9782-1_27}.

\bibitem{newey2018}
Newey WK, Robins JR.
\newblock Cross-Fitting and Fast Remainder Rates for Semiparametric Estimation.
\newblock arXiv preprint arXiv:180109138. 2018.
\newblock Available from: \url{https://arxiv.org/abs/1801.09138}.

\bibitem{chernozhukov2018}
Chernozhukov V, Chetverikov D, Demirer M, Duflo E, Hansen C, Newey W, et~al.
\newblock Double/debiased machine learning for treatment and structural parameters.
\newblock The Econometrics Journal. 2018;21(1):C1-C68.
\newblock Available from: \url{https://doi.org/10.1111/ectj.12097}.

\bibitem{foster2023}
Foster DJ, Syrgkanis V.
\newblock {Orthogonal statistical learning}.
\newblock The Annals of Statistics. 2023;51(3):879-908.
\newblock Available from: \url{https://doi.org/10.1214/23-AOS2258}.

\bibitem{bouvier2024}
Bouvier F, Peyrot E, Balendran A, S\'egalas C, Roberts I, Petit F, et~al.
\newblock Do machine learning methods lead to similar individualized treatment rules? A comparison study on real data.
\newblock Statistics in Medicine. 2024;43(11):2043-61.
\newblock Available from: \url{https://doi.org/10.1002/sim.10059}.

\bibitem{bach2024}
Bach P, Kurz MS, Chernozhukov V, Spindler M, Klaassen S.
\newblock {DoubleML}: An Object-Oriented Implementation of Double Machine Learning in {R}.
\newblock Journal of Statistical Software. 2024;108(3):1-56.
\newblock Available from: \url{https://www.jstatsoft.org/article/view/v108i03}.

\bibitem{okasa2022}
Okasa G.
\newblock Meta-Learners for Estimation of Causal Effects: Finite Sample Cross-Fit Performance.
\newblock arXiv. 2022.
\newblock ArXiv:2201.12692 [econ.EM].
\newblock Available from: \url{https://arxiv.org/abs/2201.12692}.

\bibitem{ellul2025}
Ellul S, Vansteelandt S, Carlin JB, Moreno-Betancur M.
\newblock Causal Machine Learning Methods and Use of Cross-Fitting in Settings With High-Dimensional Confounding.
\newblock Statistics in Medicine. 2025;44(20--22):e70272.
\newblock Available from: \url{https://doi.org/10.1002/sim.70272}.

\bibitem{mcclean2024}
McClean A, Balakrishnan S, Kennedy EH, Wasserman L.
\newblock Double Cross-fit Doubly Robust Estimators: Beyond Series Regression.
\newblock arXiv preprint arXiv:240315175. 2024.
\newblock Available from: \url{https://arxiv.org/abs/2403.15175}.

\bibitem{fisher2023}
Fisher A, Fisher V.
\newblock Three-way Cross-Fitting and Pseudo-Outcome Regression for Estimation of Conditional Effects and other Linear Functionals.
\newblock arXiv preprint arXiv:230607230. 2023.
\newblock Available from: \url{https://arxiv.org/abs/2306.07230}.

\bibitem{ke2024}
Ke D, Zhou X, Yang Q, Song X.
\newblock Doubly Robust Triple Cross-Fit Estimation for Causal Inference with Imaging Data.
\newblock Statistics in Biosciences. 2024.
\newblock Online first.
\newblock Available from: \url{https://doi.org/10.1007/s12561-024-09458-1}.

\bibitem{chiang2022}
Chiang HD, Kato K, Ma Y, Sasaki Y.
\newblock Multiway Cluster Robust Double/Debiased Machine Learning.
\newblock Journal of Business \& Economic Statistics. 2022;40(3):1046-56.
\newblock Available from: \url{https://doi.org/10.1080/07350015.2021.1895815}.

\bibitem{zivich2021}
Zivich PN, Breskin A.
\newblock Machine Learning for Causal Inference: On the Use of Cross-fit Estimators.
\newblock Epidemiology. 2021;32(3):393-401.
\newblock Available from: \url{https://doi.org/10.1097/EDE.0000000000001332}.

\end{thebibliography}
\end{document}